 \documentclass[apj]{emulateapj}
\usepackage{color}

 \shorttitle{Electron Shock Acceleration}
\shortauthors{Guo and Giacalone}
\singlespace
\begin{document}

\title{The Effect of Large Scale Magnetic Turbulence on the Acceleration of Electrons by Perpendicular Collisionless Shocks}

\author{Fan Guo and Joe Giacalone}
 \doublespace
\affil{Department of Planetary Sciences, University of Arizona,
Tucson, AZ 85721, USA}

\email{guofan@lpl.arizona.edu}

\begin{abstract}
We study the physics of electron acceleration at collisionless shocks that move
through a plasma containing large-scale magnetic fluctuations. We numerically
integrate the trajectories of a large number of electrons, which are treated as
test particles moving in the time dependent electric and
magnetic fields determined from 2-D hybrid simulations (kinetic ions, fluid
electron). The large-scale magnetic fluctuations effect the electrons in a
number of ways and lead to efficient and rapid energization at the shock front.
Since the electrons mainly follow along magnetic lines of force, the
large-scale braiding of field lines in space allows the fast-moving electrons
to cross the shock front several times, leading to efficient acceleration.
Ripples in the shock front occuring at various scales will also contribute to
the acceleration by mirroring the electrons. Our calculation shows that this
process favors electron acceleration at perpendicular shocks. The current study
is also helpful in understanding the injection problem for electron
acceleration by collisionless shocks. It is also shown that the spatial
distribution of energetic electrons is similar to in-situ observations
\citep[e.g.,][]{Bale1999GRL,Simnett2005}. The process may be important to our
understanding of energetic electrons in planetary bow shocks and interplanetary
shocks, and explaining herringbone structures seen in some type II solar radio
bursts.
\end{abstract}


\section{Introduction}
Collisionless shocks are widely believed to be the primary acceleration
mechanism giving rise to the ubiquitous existence of energetic particles
 in space. The theory of diffusive shock acceleration (DSA) was proposed some
 $30$ years ago \citep{Axford1978,Bell1978MNRAS,Blandford1978,Krymsky1977} and
 is currently believed to be the most important mechanism for a variety of
  astrophysical environments, for example,
interplanetary shocks, the heliospheric termination shock, and
shocks associated with supernova remnants. The theory predicts a
universal power-law energy flux spectrum $dJ/dE\propto E^{-1}$ for
strong shocks with a density compression ratio of 4.

The physical mechanism by which particles are accelerated from thermal energies
 to much higher energies where DSA is presumed to be applicable (the injection
  problem) has received some recent attentions but has not
reached a common consensus explanation. Many acceleration theories,
for example, shock drift acceleration (SDA) \citep[see
reviews,][]{Armstrong1985,Decker1988SSRv} and shock surfing
acceleration \citep{Sagdeev1966RvPP,Lee1996JGR,Zank1996JGR} have
been proposed. The acceleration of low energy protons in the shocks
containing large-scale pre-existing magnetic fluctuations is very
efficient \citep{Giacalone2005ApJa,Giacalone2005ApJb}, which
suggests that there may not be an injection problem.

It is generally thought that,
 pre-accelerated particles will interact resonantly with magnetic
  turbulence which results in isotropization and diffusion.
    Many previous works considering magnetic turbulence focused on the ions
\citep[e.g.,][]{Bell1978MNRAS,Giacalone1992GeoRL,Ng2003ApJ,Giacalone2004ApJ,Giacalone2005ApJa,Giacalone2005ApJb}.
However, the acceleration of electrons is less well understood since
for electrons whose gyroradii are very small, the cyclotron
 resonance condition is
not easily satisfied thus they cannot interact resonantly with
large-scale ambient turbulence on ion-scale. While the scattering
provided by whistler waves \citep{Shimada1999} is one possibility,
\citet{Jokipii2007ApJ} proposed an attractive solution to the
injection problem that does not require pitch-angle scattering,
i.e., conserving the first adiabatic invariant. The idea is that the
low-rigidity particles, especially electrons, can move rapidly along
meandering magnetic field lines and thus travel back and forth
between shock front. The particles gain energy from the difference
between upstream and downstream flow velocities.


Energetic electrons are often observed to be associated with
collisionless shocks. Accelerated electrons are thought to produce
type II radio bursts in the solar corona and interplanetary space.
\citet{Anderson1979GeoRL} reported \emph{ISEE} spacecraft
measurements of upstream electrons ($>16$ keV) of the Earth's bow
shock that originate from a thin region close to the point of
tangency between interplanetary magnetic field lines and the shock
surface. \citet{Tsurutani1985JGR} showed observations of energetic
electrons associated with interplanetary shocks showing "spike-like"
flux enhancements for energies $\gtrsim 2$ keV. The spike events
were observed at quasi-perpendicular shocks with $\theta_{Bn}
\gtrsim 70^\circ$, where $\theta_{Bn}$ is the angle between upstream
magnetic field and shock normal. Some shock crossings had no
enhancements of energetic electrons which were reported to be
associated with low shock speeds and small $\theta_{Bn}$.
\citet{Simnett2005} presented data which shows energetic electrons
are accelerated close to shock front. They also showed some
accelerated electrons can escape far upstream of quasi-perpendicular
interplanetary shocks. The clear evidence of electron acceleration
by DSA is rare, but a recent example was discussed by
\citet{Shimada1999} showing the importance of whistler waves.

In order to explain the energization of electrons within the shock
layer, \citet{WuCS1984} and \citet{Leroy1984} developed analytic
models for electron acceleration from thermal energies by adiabatic
reflection by a quasi-perpendicular shock. This is known as
fast-Fermi acceleration. This theory describes a scatter-free
electron acceleration process in a planar, time-steady shock. It
obtains a qualitative agreement with observations at Earth's bow
shock in terms of the loss-cone pitch-angle distribution and energy
range of accelerated electrons. \citet{Krauss-Varban1989a} used the
combination of electron test particle simulation and 1-D hybrid
simulation and verified Wu's basic conclusions. The main energy
source of fast Fermi acceleration comes from the $-\textbf{V} \times
\textbf{B}/c$ electric field which is the same as SDA
\citep{Armstrong1985}. It can also be demonstrated that fast-Fermi
acceleration and SDA are the same process in two different frames of
reference \citep{Krauss-Varban1989JGRb}. Thus one would expect
electrons to drift in the direction perpendicular to the flow and
magnetic field.
For a single reflection, the fraction and energies of accelerated
particles are limited \citep[e.g.,][]{Ball2001PASA}.
\citet{Holman1983ApJ} proposed the basic outline for type II solar
radio bursts in which energetic electrons are accelerated through
SDA. It is expected that multiple reflections are required in order
to explain herringbone structures in type II bursts, where the
electrons are accelerated to a fraction of the speed of light. More
recently, \citet{Burgess2006} studied electron acceleration in 2-D
quasi-perpendicular shocks using test-particle simulations and
self-consistent hybrid simulations. He focused on the effect of the
rippling of the shock front on particle acceleration in highly
oblique shocks with $\theta_{Bn} \geq 80^\circ$. The ripples, in
this case, were produced by instabilities along the shock front.
Burgess found that the acceleration of electron by SDA can be more
efficient for a rippled shock. The shape of the resulting energy
spectra has a flat plateau from the initial release energies to
energies several times higher than this. Above the flat plateau the
spectra drop off steeply as $\theta_{Bn}$ become smaller.

In this paper, we use test particle simulations combined with $2$-D
hybrid simulations that include pre-existing large-scale magnetic
field turbulence \citep{Giacalone2005ApJb} to study the shock
acceleration of electrons. In addition to the effect of the
large-scale turbulence, the shock microphysics occuring on ion
length and time scales is also included. Moreover, the shock front
is rippled and distorted in response to the turbulence, which is
also included in our model. In $\S $ 2 we describe the numerical
method we used to combine the fields from the hybrid simulation and
test-particle simulation to obtain the electron distribution. $\S $
3 gives the main results of our simulation. In $\S $ 4 we summarize
the main conclusions and discuss the implication of our work.

\section{Numerical Method}
Investigating particle transport in the vicinity of a collisionless
shock requires a spatial scale large enough for particles to
propagate back and forth across the shock, and a spatial resolution
small enough to include the detailed physics for particle scattering
and shock microstructure. We implement a combination of a $2$-D
hybrid simulation to model the fields and plasma flow and a test
particle simulation to follow the orbits of a large number of
energetic electrons. In the first step, we employ a two-dimensional
hybrid simulation similar to previous work \citep{Giacalone2005ApJb}
that includes pre-existing large-scale turbulence. In the hybrid
simulation \citep[e.g.,][]{Winske1988JGR}, the ions are treated
fully kinetically and thermal (i.e., non-energetic) electrons are
treated as a massless fluid. This approach is well suited to resolve
ion-scale plasma physics which is critical to describe supercritical
collisionless shocks. In this study, we consider a two-dimensional
Cartesian grid in the $x-z$ plane. All the physical vector
quantities have components in three directions, but depend spatially
only on these two variables. A shock is produced by using the
so-called piston method \citep[for a discussion,
see][]{Jones1991SSRv}, in which the plasma is injected continuously
from one end ($x=0$, in our case) of the simulation box, and
reflected elastically at the other end ($x=L_x$). This boundary is
also assumed to be a perfectly conducting barrier. The pileup of
density and magnetic field creates a shock propagating in the $-x$
direction. To include the large-scale magnetic fluctuations, a
random magnetic field is superposed on a mean field at the beginning
of the simulation and is also injected continuously at the $x=0$
boundary during the simulation. The simplified one-dimensional
fluctuations have the form $\textbf{B}(z, t) = \delta \textbf{B}(z,
t) + \textbf{B}_1$, where $\textbf{B}_1$ is the averaged upstream
magnetic field. The fluctuating component contains an equal mixture
of right- and left-hand circularly polarized, forward and backward
parallel-propagating plane Alfven waves. The amplitude of the
fluctuations is determined from a Kolmogorov power spectrum:

$$
P(k) \propto \frac{1}{1 + (k L_c)^{5/3}}
$$

\noindent in which $L_c$ is the coherence scale of the fluctuations, see
\citep{Giacalone2005ApJb} for more details. For the simulations presented in
this study, we take $L_c = L_z$, which is the size of simulation box in z
direction. Note that in addition to magnetic fluctuations,
there are also velocity perturbations with $\delta \textbf{v} = v_{A1}\delta
\textbf{B}/B_1$ (Alfven waves). For most of the parts in the paper, we consider
a turbulence variance $\sigma = \delta B^2/B_1^2 = \delta v^2/v_{A1}^2 = 0.3$,
where $\delta v$ and $v_{A1}$ are the magnitude of velocity perturbation and
upstream Alfven speed, respectively. We also discuss the effect of different
values of turbulence variances.

The size of the simulation box $L_x\times L_z = 400 c/\omega_{pi}\times 1024
c/\omega_{pi}$, where $c/\omega_{pi}$ is the ion inertial length. The Mach
number of the flow in the simulation frame is $M_{A0} = 4.0$, the averaged Mach
number in the shock frame is about $5.6$. Most of the results presented here
are for averaged shock normal angle $<\theta_{Bn}> = 90^\circ$, but we also
simulate the cases for $<\theta_{Bn}> = 60^\circ$ and $75^\circ$ to examine the
dependence of the acceleration efficiency on shock normal angle. The other
important simulation parameters include electron and ion plasma beta $\beta_e =
0.5$ and $\beta_i = 0.5$, respectively, grid sizes $\Delta x = \Delta z = 0.5
c/\omega_{pi}$, time step $\Delta t = 0.01 \Omega_{ci}^{-1}$,
the ratio between light speed and upstream Alfven speed
$c/v_{A1} = 8696.0$, and the anomalous resistivity is taken to be $\eta = 1
\times 10^{-5} 4\pi\omega_{pi}^{-1}$. The initial spatially uniform thermal ion
distribution was generated using $40$ particles per cell. Different from
previous studies, the consideration of large-scale magnetic fluctuations
enables us consider the effect of pre-existing magnetic turbulence on electron
acceleration, which has been shown to be important for low-energy ion
acceleration \citep{Giacalone2005ApJa,Giacalone2005ApJb} since particle
transport normal to the mean field is enhanced. However, the particle transport
in full $3$-D turbulence can not be properly treated in a self-consistent way
using available computation. As demonstrated by previous works
\citep{Jokipii1993,Giacalone1994ApJ,Jones1998ApJ}, in the model with at least
one ignorable coordinate, the center of gyration of particles is confined to
within one gyroradius of the original magnetic field line. The test-electrons
can still move normal to the mean field in our model because of the field-line
random walk.

In the second part of our calculation we integrate the full motion equation of
an ensemble of test-particle electrons in the electric and magnetic fields
obtained in the hybrid simulations (see Figure $1$). This part
of the calculation is done separately from the main hybrid simulation as a post
processing phase. We assume non-relativistic motion which is reasonable because
the highest energy electrons obtained in our study are still non-relativistic.
As noted by \citet{Krauss-Varban1989a}, high-order interpolation of fields is
required to ensure numerical accuracy and avoid artificial scattering in
calculating electron trajectories. In this work we use second-order spatial
interpolation and linear temporal interpolation, which ensure
the smooth variations of the electromagnetic fields. We release a shell
distribution of electrons with energy of $100$ eV, which
corresponds to an electron velocity $V_e = 30.7 U_1 = 5.7 v_{the}$ in the
upstream frame, where $U_1$ is upstream bulk velocity in the shock frame and
$v_{the}$ is the thermal velocity of fluid electrons considered in the hybrid
simulations, respectively. This energy is typical for the halo component of
electron velocity distributions observed in solar wind. The test-particle
electrons are released uniformly upstream at $t = 70\Omega_{ci}^{-1}$ when the
shock has fully formed and is far from the boundaries. The numerical technique
used to integrate electron trajectories is the so-called Bulirsh-Stoer method,
which is described in detail by \citet{Press1986}. It is highly accurate and
conserves energy well. It is fast when fields are smooth compared with the
electron gyroradius. The algorithm uses an adjustable time-step method based on
the evaluation of the local truncation error. The time step is allowed to vary
between $5 \times 10^{-5} $ and $0.1 \Omega_{ce}^{-1}$, where $\Omega_{ce}$ is
the electron gyrofrequency. The ratio $\Omega_{ce}/\Omega_{ci}$ is taken to be
the realistic value $1836$. The total number of electrons in the simulation is
$1.6 \times 10^6$. The electrons which reach the left or right boundary are
assumed to escape from the shock region and are removed from simulation. The
boundary condition in the $z$ direction is taken to be periodic. The readers
are referred to \citep{Burgess2006} for more details on the numerical methods.

Magnetic field turbulence has already been proved to have key effect
on the particle acceleration in collisionless shocks. Unfortunately,
solving the whole problem in three-dimensional space and resolving
magnetic turbulence from coherence scale to electron scale are still
limited by available computation in the near future. This limitation
motivates us to solve these problems approximately. We also note
that in our model the electron test particle simulation is not
self-consistent since the hybrid simulation does not include the
electron scale physics. The electron scale shock structure, which
may be important is neglected here.

\section{Simulation Results}
Figure $1$ shows a snapshot of the $z$ component of the magnetic field,
$B_z/B_1$, at time $110 \Omega^{-1}_{ci}$ in a gray-scale representation,
overlayed by a two-dimensional magnetic vector field. At this
time, the shock is fully developed. In this case, the angle between the average
magnetic field direction and shock normal, $<\theta_{Bn}>$ is $90^\circ$. The
position of the shock front is clearly seen from the boundary of the magnetic
field jump. The shock is moving in the $-x$ direction at a
speed dependent on $z$, which is about 1.6 $v_{A1}$ on average. Because of the
effect of large-scale turbulence with the shock, the shock surface become
irregular on a variety of spatial scales from small-scale ripples, which could
be due to ion-scale plasma instabilities \citep{Lowe2003}, to large-scale
structure caused by the interaction between the shock and upstream turbulence
\citep{Neugebauer2005,Giacalone2008ApJ,Lu2009ApJ}. The upstream magnetic field
is compressed and distorted as it passes through the shock into the downstream
region. We note that the rippling of the shock and varying upstream magnetic
field leads to a varying local shock normal angle along the shock front. As we
will discuss later, the irregular shock surface and magnetic field geometry
will efficiently accelerate electrons and produce a number of features similar
to observations, such as the electron foreshock and spike-like intensity
increases at the shock front. The meandering of field lines close to the shock
surface helps to trap the electrons at the shock, leading to efficient
acceleration. The shock ripples also contribute to the acceleration by
mirroring electrons between them.

Figure $2$ shows a color-coded representation of the number of energetic
electrons with energies higher than $10$ times (i.e., $1$ keV) the initial (at
release) energy at three different times (a) $76\Omega_{ci}^{-1}$, (b)
$81\Omega_{ci}^{-1}$, (c) $90\Omega_{ci}^{-1}$, respectively. It is found that
after the initial release, a fraction of the electrons are reflected and
accelerated at the shock front, and then travel upstream along the turbulent
magnetic field lines. These accelerated electrons are then taken back to the
shock by the field line meandering, which provides even further acceleration.
The number of energetic electrons close to the shock surface is highly
irregular because the acceleration efficiency varies along the shock front
depending on the local shock normal angle \citep{WuCS1984}. Most of the
electrons are concentrated near the shock front since the global magnetic field
is mostly perpendicular to the shock normal. As the field lines convect through
shock, the electrons eventually are taken downstream. Since the electrons are
tied to individual field lines in 2-D magnetic field, once the electrons are no
longer capable of crossing the shock, there will be no additional significant
acceleration. At this point, once all electrons are downstream, the energy
spectrum no longer changes with time.

Examination of the trajectories of some electrons shows that the rippling of
the shock front also contributes to the acceleration by mirroring electrons
between the ripples, as illustrated in Figure $3$. In this figure, the top left
plot displays the trajectory of a representative electron in the $x-z$ plane,
overlapped with the 2-D gray-scale representation of $B_z$ at $\Omega_{ci}t =
89.0$, the gray scale is the same as in Figure $1$. The upper right plot shows
the position of this electron (in $x$) as a function of time. The electron
bounces back and forth between the ripples for several times. For example, the
reflections are labeled $a-b$, $c$, $d$, $f-g$, $h$, and $j$. The energy change
as a function of position, $x$, corresponding to these reflections is shown in
the bottom left panel. We find that there are jumps in energy at each of the
reflections. The panel on the bottom right shows the electron energy as a
function of time which also illustrates the features of multiple accelerations
related to multiple reflections. The trajectory analysis shows the electron
will be mirrored between the ripples for a couple of times and by this get
accelerated multiple times. Note that the shock does not move
much during the time scale of this trajectory.

We now consider the effect of varying the angle between the mean magnetic field
and shock-normal. Shown in Figure $4$ are the resulting energy spectra for
three different mean shock-normal angles ($<\theta_{Bn}>= 60^\circ, 75^\circ$
and $90^\circ$, respectively) at the end of simulations ($\Omega_{ci}t=120.0$).
It is found that for $<\theta_{Bn}> = 90^\circ$, the electrons can readily be
accelerated to up to $200-300$ times the initial energy within $50
\Omega_{ci}^{-1}$. The spectrum is flat between about $0.1$ keV to $0.7$ keV.
This shape is similar to the "plateau" structure discussed by
\citet{Burgess2006}. Above $1$ keV, the spectrum falls off with energy as a
slope index about $-3$. It can be found that both the number fraction and
highest energy of accelerated particles decreases as $<\theta_{Bn}>$ decreases.
We have also tried different value of initial energies (not
shown), and find that the acceleration efficiency decreases for electrons with
higher initial energies, which is similar to the results of
\citep{Burgess2006}.

The effect of different strengthes of magnetic turbulence is
examined in Figure $5$. We compare three cases with different turbulence
variances $\sigma = 0.1$, $0.3$, and $0.5$, respectively. At the end of
simulations, the final energy spectra are similar at low energies, with
significant variations in the spectra only at energies larger than $2$ keV. It
is found that the energy spectrum is hardened at high energies when the
turbulence variance is largest, which indicates the large-scale turbulence is
more important for accelerating electrons to high energies. We argue that
collisionless shocks which move through magnetic turbulence with significant
power leads to efficient electron acceleration to high energies since the
motion normal to the shock front is enhanced. The reason is that the meandering
of field lines is enhanced, which allows the electrons have a better chance to
travel though the shock multiple times.

We note that the spatial distribution of energetic electrons is determined not
only by the ripples in the shock front, but also by the global topology of the
magnetic field lines. An example is shown in Figure $6$, which shows the
profiles of the number of energetic electrons at $\Omega_{ci}t=100.0$ as a
function of $x$, for the case of $<\theta_{Bn}>=90^\circ$. The black solid line
is the profile at $z = 200 c/\omega_{pi}$, and the red dash line shows the
profile at $z = 800 c/\omega_{pi}$. The corresponding position of the shock
front at each of these values of $z$ are represented using dot lines. At $ z =
200 c/\omega_{pi}$, it is observed that the energetic electrons travel far
upstream up to about $100 c/\omega_{pi}$. However, the profile at $z = 800
c/\omega_{pi}$ shows no significant upstream energetic electron flux. The
upstream energetic electron profiles show irregular features similar to in-situ
observations reported by \citet{Simnett2005} (in Figure 10). The irregular
features are controlled by the global topology of the large-scale turbulent
magnetic field lines, along which the accelerated electrons could travel far
upstream. Additionally, energetic electron profiles in $x$ direction generally
show "spike-like" structure close to the shock front, which is usually observed
in interplanetary shocks and Earth's bow shock. We note this
feature is relatively stable within the simulation time once the upstream
electron structure developed.

\section{Discussion and Conclusions}
We studied the acceleration of electrons at collisionless shocks by
utilizing a combination of a 2-D hybrid simulation to obtain the
shock structure and a test-particle simulation to determine the
motion of electrons. The hybrid simulation provides realistic
electric and magnetic fields within the transition layer of the
shock that effect the motion of test-electrons, which is determined
by solving the equation of motion. The interaction of the shock with
pre-existing upstream fluctuations, and other nonlinear processes
occuring in the hybrid simulation lead to a "rippling" of shock
surface which also effects the transport of the electrons.

We find that the electrons are efficiently accelerated by a nearly
perpendicular shock.  The turbulent magnetic field, leads to field-line
meandering which allows the electrons to cross the shock front many times. The
rippling of the shock front also contributes to the acceleration by mirroring
electrons between the ripples. For the case that the averaged shock normal
angle $<\theta_{Bn}> = 90^\circ$ and turbulence variance $\sigma = 0.3$, the
electrons can readily be accelerated to up to $200-300$ times the initial
energy. The resulting spectrum is flat between about $0.1$ keV to $0.7$ keV. At
higher energies, the spectrum falls off with energy like a power law with a
spectral slope of about $-3$. This acceleration process is more efficient at
perpendicular shocks. As $<\theta_{Bn}>$ decreases from $90^\circ$ , both the
number fraction and highest achievable energy of accelerated particles
decreases. Based on our calculations, we conclude that perpendicular shocks are
the most important for the acceleration of electrons. The current study is
helpful in understanding the injection problem for electron acceleration by
collisionless shocks. It is also found that different value of
magnetic turbulence variances strongly affects the maximum energy attainable.
The case with larger turbulence variance has a flatter energy spectrum than the
case of smaller turbulence variance, which suggests the enhanced motion of
electrons normal to the shock front, due to enhanced field-line random walk, is
of importance for the acceleration of electrons to high energies.

In addition, we also found that the energetic electron density
upstream and downstream of collisionless shocks show filamentary
structures (Figure 2). This could help explain electron spike-like
events observed upstream and downstream of terrestrial and
interplanetary shocks
\citep{Anderson1979GeoRL,Tsurutani1985JGR,Simnett2005}. Observation
by Voyager $1$ at the termination shock and in the heliosheath also
show the evidence of electron spike-like enhancements at the shock
front \citep{Decker2005Sci}. The upstream spatial distribution of
energetic electrons shows irregular features which depend on both
the irregularity in the shock surface and the global topology of
magnetic field lines. At first the electrons are accelerated and
reflected at the shock front, and then they travel upstream along
the magnetic field lines. The electrons could be taken far upstream
by field line random walk. This result can possibly lead to an
interpretation to the complex electron foreshock events recently
observed to be associated with interplanetary shocks
\citep{Bale1999GRL,Pulupa2008ApJ}.  \citet{Bale1999GRL} and
\citet{Pulupa2008ApJ} proposed the complex upstream electron events
are resulted from large-scale irregularities in shock surface. In
this paper we have demonstrated that the upstream electron flux may
be controlled by both an irregular shock surface and large-scale
meandering magnetic field lines.

\section*{Acknowledgement}
The authors would like to thank Dr. D. Burgess for sharing the
details of numerical methods and Dr. J. R. Jokipii for valuable
discussions. This work was supported in part by NSF under grant
ATM0447354 and by NASA under grant NNX07AH19G.

\clearpage
\begin{figure}
 \begin{center}
\includegraphics[width=35pc]{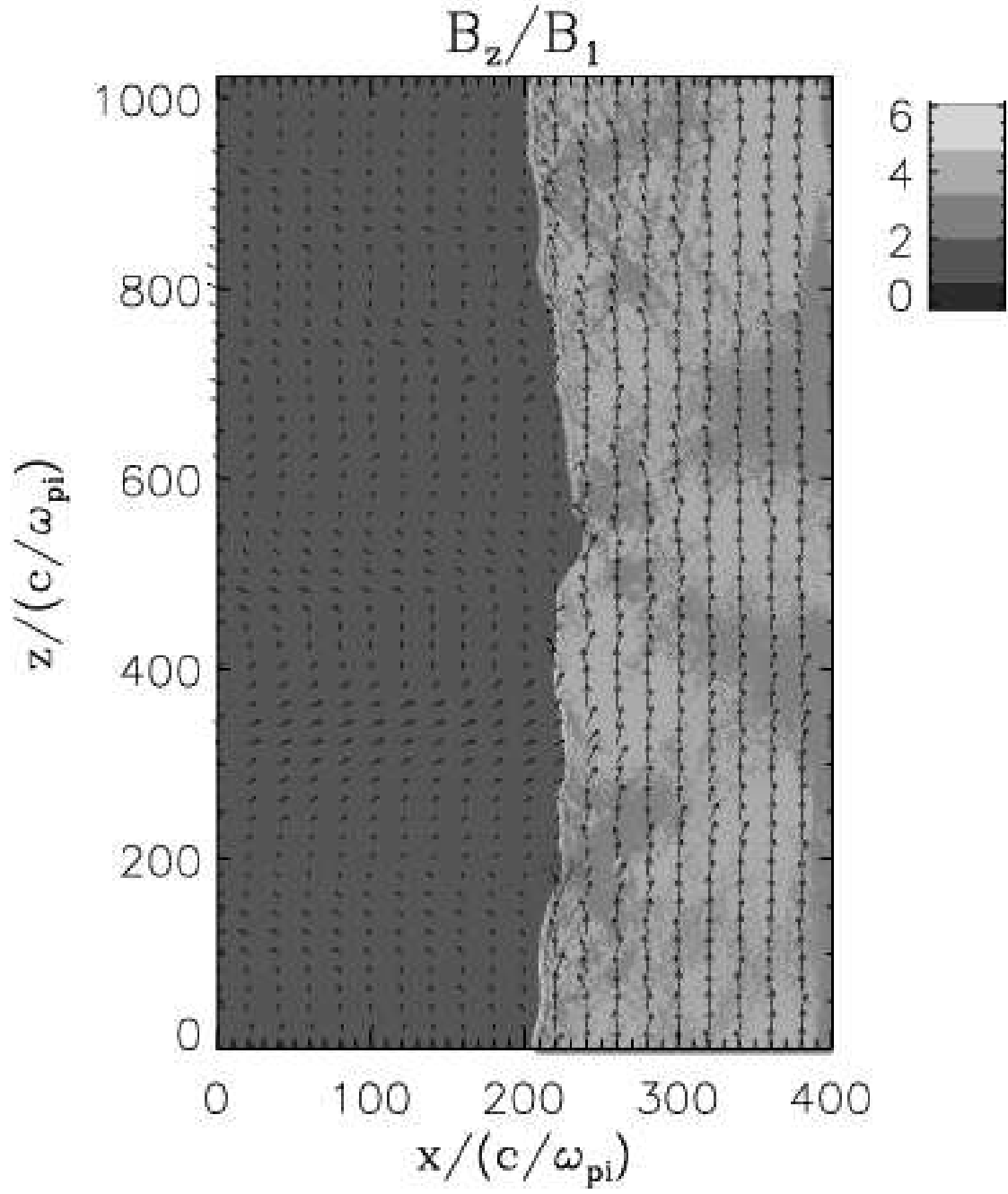}
 \caption{A snapshot of magnetic field in $z$ direction $B_z/B_1$ represented in gray-scale at time $110\Omega_{ci}^{-1}$,
 where $B_1$ is the averaged upstream magnetic field strength. A
 two-dimensional vector field is also overlapped, which indicates the direction of magnetic field.
 The shock surface is shown to be rippled and irregular in different scales. }
 \end{center}
 \end{figure}

 \begin{figure}
 \begin{center}
\includegraphics[width=30pc,angle=90,scale=1.5]{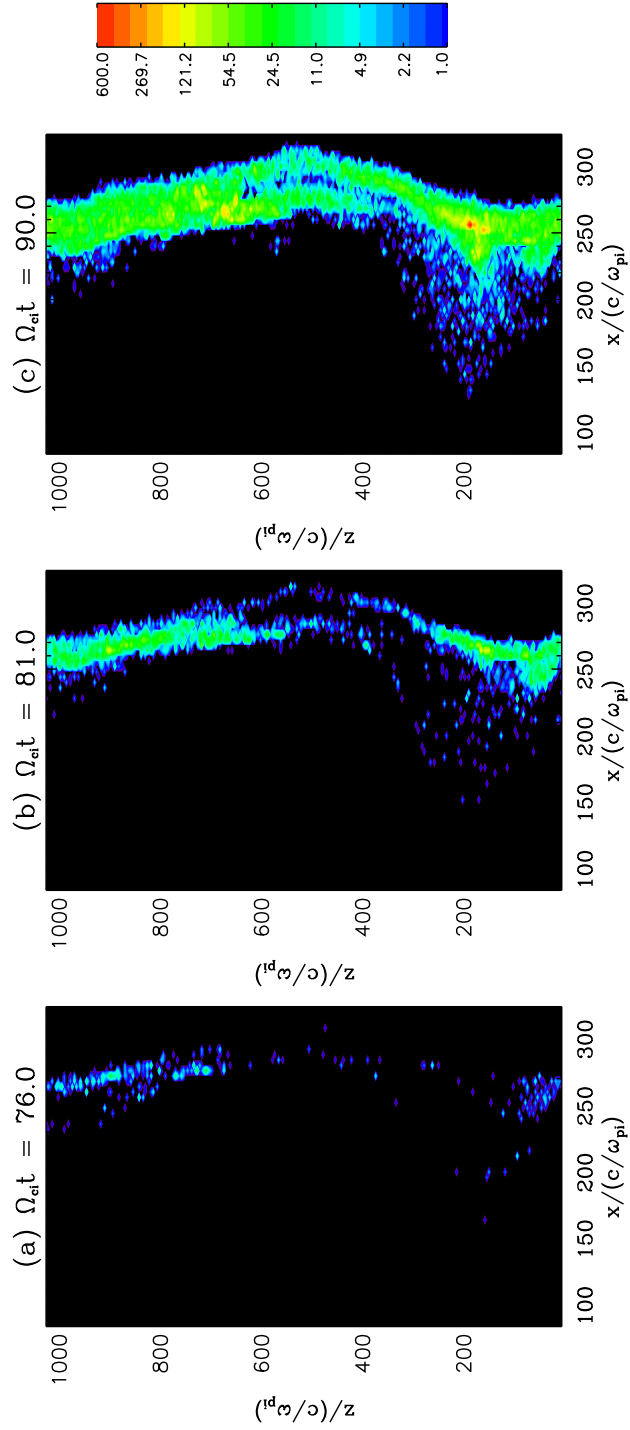}
 \caption{The number of energetic electrons with energies $E>10E_0$, the initial release energy $E_0 = 100$
 eV, at (a) $\Omega_{ci}t=76$, (b) $\Omega_{ci}t=81$, and (c)
 $\Omega_{ci}t=90$, respectively. Initially electrons are released uniformly upstream at $\Omega_{ci}t=70$.}
 \end{center}
 \end{figure}

 \begin{figure}
 \begin{center}
\includegraphics[width=30pc]{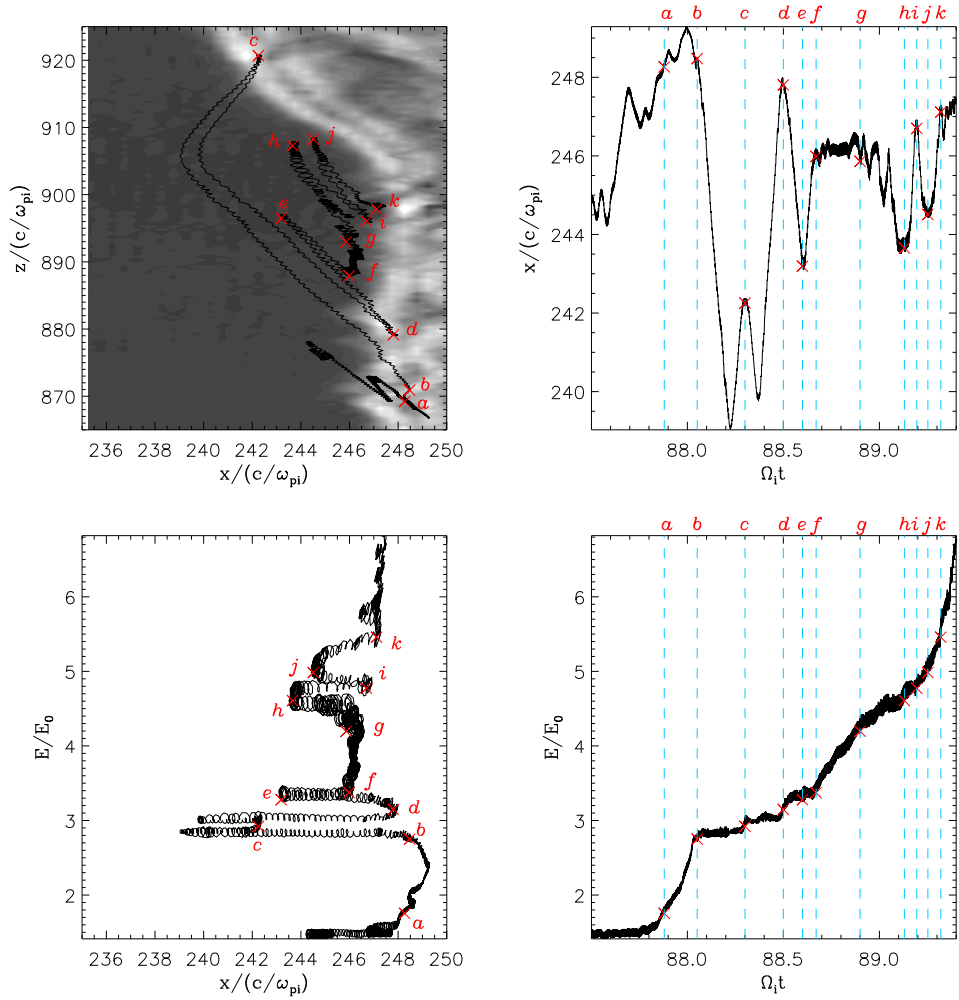}
 \caption{A typical electron trajectory analysis which shows acceleration by multiple mirroring between ripples.
 The top left panel displays the trajectory of the representative electron in $x-z$ plane, overlapped with contour
 of $B_z$ magnetic field where the gray-scale is the same as that in Figure 1; The top right panel shows the
 position of the electron in $x$ coordinate as a function of time; The bottom left panel illustrates the energy of
 the representative electron $E/E_0$ as a function of $x$; The bottom right panel shows the dependence of electron energy
$E/E_0$ on time.}
 \end{center}
 \end{figure}

 \begin{figure}
 \begin{center}
\includegraphics[width=25pc]{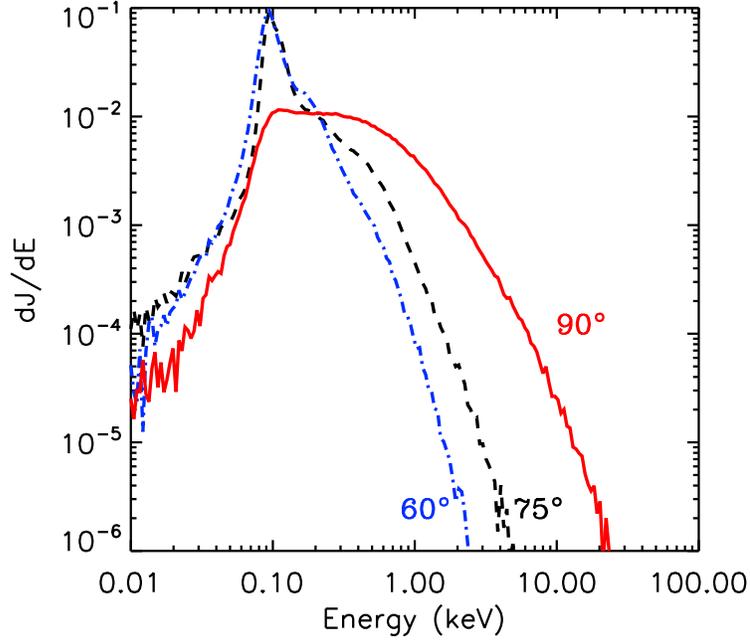}
 \caption{The energy flux spectrum of electrons at $\Omega_{ci}t=120$ for different averaged shock normal angle. The red solid line is
  in the case that the shock angle $<\theta_{Bn}>=90^\circ$, the blue dot dashed line and the black dashed line are
   in the cases that $<\theta_{Bn}> = 60^\circ $ and $ 75^\circ $, respectively.}
 \end{center}
 \end{figure}

\begin{figure}
 \begin{center}
\includegraphics[width=25pc]{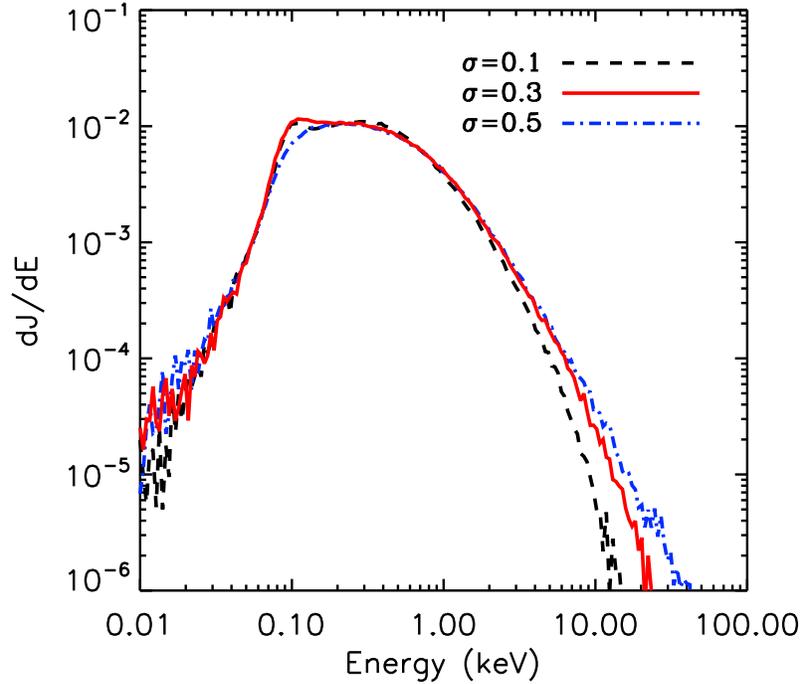}
 \caption{The energy flux spectrum of electrons at $\Omega_{ci}t=120$ for averaged perpendicular shock with different turbulence
  variances. The black dashed line, red solid line, and blue dot dashed line are in the cases that $\sigma = 0.1$, $0.3$, and $0.5$, respectively.}
 \end{center}
 \end{figure}

\begin{figure}
 \begin{center}
\includegraphics[width=25pc]{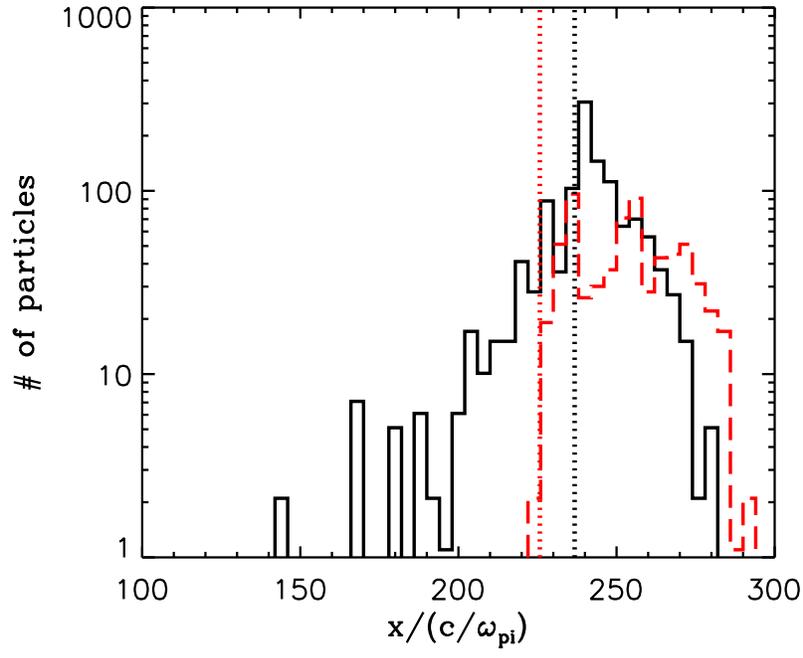}
 \caption{The solid black line and dashed red line show the profiles of the number of energetic
 electrons at $z = 200c/\omega_{pi}$ and $z = 800c/\omega_{pi}$ at time $\Omega_{ci}t=100$, respectively.
 The red dot line and black dot line label the corresponding positions of the shock fronts.}
 \end{center}
 \end{figure}

\end{document}